\documentclass[aps,twocolumn]{revtex4-2}

\usepackage{amsmath,amssymb,amsfonts,graphicx,tabularx,xcolor}

\usepackage[unicode=true,colorlinks=true]{hyperref}

\hypersetup{linkcolor=blue,citecolor=blue,urlcolor=blue}

\begin{document}

\title{Non-Abelian topological order with SO(5)$_{1}$ chiral edge states}

\author{Ying-Hai Wu}
\email{yinghaiwu88@hust.edu.cn}
\affiliation{School of Physics and Wuhan National High Magnetic Field Center, Huazhong University of Science and Technology, Wuhan 430074, China}

\author{Hong-Hao Tu}
\email{hong-hao.tu@tu-dresden.de}
\affiliation{Institut f\"ur Theoretische Physik, Technische Universit\"at Dresden, 01062 Dresden, Germany}

\begin{abstract}
We consider a chiral spin liquid constructed using the parton theory. This state supports non-Abelian anyons and neutral fermions, which share some similarities with the celebrated Moore-Read state. The edge physics of the parton state and the Moore-Read state is very different. Based on conformal field theory (CFT) analysis, it is proposed that the edge states exhibit an emergent SO(5) symmetry. The counting of edge states in the low-lying SO(5)$_{1}$ CFT towers is computed. The chiral spin liquid is generated using tensor network methods, which allows us to confirm the SO(5)$_{1}$ counting using entanglement spectrum. An additional feature of multiple branches in the entanglement spectrum is observed and analyzed.
\end{abstract}

\maketitle

\section{Introduction}

The existence of identical particles underlies the great wonders in the universe. After the establishment of quantum mechanics, it was realized that a many-body wave function may display two different behaviors when two particles are exchanged. The wave function remains intact if the exchanged particles are bosons, but acquires a minus sign if they are fermions. Half a century later, Leinaas and Myrheim proposed to interpret the exchange process as continuous and introduced the concept of ``braid statistics"~\cite{Leinaas1977}. In two dimensions, Artin's braid group makes it possible to have certain objects that are neither bosons nor fermions, which hence acquired the name anyons~\cite{Wilczek1982,WuYS1984-1}. This interesting idea becomes experimentally relevant with the discovery of fractional quantum Hall (FQH) effect~\cite{Tsui1982,Laughlin1983,Arovas1984}. It is believed that most FQH states known to date support anyons with Abelian braid statistics where the exchange results in a nontrivial phase. This is difficult to confirm directly in experiments but very positive results have been reported~\cite{Bartolomei2020,Nakamura2020}.

An even more exotic possibility is that the ground state of a system with multiple anyons at fixed positions is degenerate. The braiding of anyons would result in a unitary transformation in the degenerate ground-state subspace and two such actions may not commute with each other. This type of anyons has been dubbed as non-Abelian and could be utilized to perform topological quantum computation~\cite{Kitaev2003,Nayak2008}. In the context of FQH states, two approaches have been adopted to construct non-Abelian anyons: one based on conformal field theory (CFT)~\cite{Moore1991,Read1999} and another based on parton theory~\cite{Jain1989-2,WenXG1991-1}. It has been shown that some states generated by these approaches are good candidates for certain experiments~\cite{Willett1987,Morf1998,Rezayi2000,XiaJS2004,Rezayi2009,Apalkov2011,Papic2014,KiDK2014,KimYW2015,Falson2015,Zibrov2017,LiJIA2017,WuYH2017-1,KimYW2019,Faugno2019,Balram2020-1}. The two approaches have also been extended to other cases beyond FQH states~\cite{Savary2016,ZhouY2017}.

The presence of anyons leads to some interesting properties that are unified in the framework of topological order~\cite{WenXG-Book}. One signature of anyons is that there are multiple degenerate ground states if the system is defined on a closed manifold with nonzero genus. This can be understood using the virtual process in which two anyons are created from the vacuum, travel along the incontractible loops of the manifold, and finally reannihilate with each other~\cite{WenXG1990}. Another intriguing aspect of topological order is nontrivial features of quantum entanglement. The most commonly used measure is the von Neumann entanglement entropy associated with certain bipartitions of the system. Its subleading term is a constant determined by the total quantum dimension of the anyons~\cite{Kitaev2006-1,Levin2006}. For a manifold with nonzero genus, the system can be divided into two parts using its incontractible loops and the entropy of any state can be computed. For the ground state subspace, we can find a particular basis that minimizes such entropy. This provides us the minimally entangled states (MES) from which braid statistics of the anyons can be extracted directly~\cite{Keski1993,ZhangY2012,ZhuW2014}.

If a topologically ordered state is placed on an open manifold, there may exist gapless modes in the vicinity of the edge. The anyon content in the bulk and the edge physics are intimately connected and generally referred to as bulk-boundary correspondence. The tunneling into gapless edge states provides useful information about the topological order~\cite{WenXG1995}. The edge physics and quantum entanglement in the bulk are related via entanglement spectrum~\cite{LiH2008}. The CFT approach is natural for exploring the bulk-boundary correspondence because the wave function for the bulk is the correlator in a CFT that describes the edge physics. In contrast, the edge physics of parton states is not as transparent and as well-understood.

The parton construction is based on the premise that each physical degree of freedom (boson, fermion, or spin) can be decomposed into two or more fictitious particles called partons. This comes at the price that the partons are not really independent but should obey certain constraints that often endow the system with some emergent gauge fields. Although the original constituents are strongly interacting, the partons are taken to be free so they form some simple product states in certain basis. In the low-energy effective field theory description, the partons may be integrated out and the gauge fields are kept. It is reasonable to expect that the edge physics can be understood using the partons, but the constraints on them are not always easy to address and could sometimes obscure the edge physics.

In this paper, we consider arguably one of the simplest non-Abelian topological states provided by the parton framework. The original proposal was an FQH state of bosons that is a product of two integer quantum Hall (IQH) states of fermions at filling factor $2$~\cite{Jain1989-2,WenXG1991-1}. This state has been recasted into a chiral spin liquid on lattice and its modular matrices and momentum polarization have been studied~\cite{ZhangY2013,ZhangY2014}. The anyon content of this state is similar to that of the Moore-Read state~\cite{Moore1991,Kitaev2006-2,Greiter2009,Glasser2015,Wildeboer2016,Lecheminant2017,LiuZX2018,ChenJY2018,ZhangHC2021}, but its edge physics has not been completely elucidated and we aim to do so in this paper. In Sec.~\ref{edge}, conformal field theory is employed to analyze the edge physics. In Sec.~\ref{liquid}, we perform numerical calculations using matrix product states (MPS) to support our theoretical predictions. The paper is concluded with some outlook in Sec.~\ref{conclusion}.

\section{SO(5)$_{1}$ Edge Theory}
\label{edge}

It is convenient to begin our discussion with continuum Landau levels (LLs). The IQH state in which the lowest $n$ LLs are completely filled is denoted as $\chi_{n}(\{z_{j}\})$ with $z_{j}$ being the complex coordinates. To construct bosonic FQH states, one boson is decomposed into two fermionic partons that form their respective IQH states with filling factors $n_{1}$ and $n_{2}$. The simple choice $n_{1}=n_{2}=2$ lead to a non-Abelian state with wave function $\Psi(\{z_{j}\})=\chi_{2}(\{z_{j}\})\chi_{2}(\{z_{j}\})$. The partons has an SU(2) symmetry that requires the introduction of an emergent gauge field. If the parton fields are integrated out, one obtains a low-energy effective field theory for this state in terms of the gauge fields~\cite{WenXG1991-1,ZhangY2013,ZhangY2014}. The edge physics is captured by U(4)$_{1}$/SU(2)$_{2}$ (a quotient of two CFTs), where U(4)$_{1}$ comes from the chiral Dirac fermions of the parton edge modes (central charge $c=4$), and SU(2)$_{2}$ is due to the gauge symmetry~\cite{ZhangY2014}. We note that SU(2)$_{2}$ has a free field representation in terms of three Majorana fermions (central charge $c=3/2$) and U(4)$_{1}$ can be interpreted as eight Majorana fermions. This means that the coset U(4)$_{1}$/SU(2)$_{2}$ is nothing but the SO(5)$_{1}$ Wess-Zumino-Witten (WZW) model, which is a free theory of five Majorana fermions ($c = 4-3/2 = 5/2$).

To better understand the edge theory, it is helpful to sketch some results about the chiral SO(5)$_{1}$ WZW model~\cite{Francesco1997}. If the system is placed on a one-dimensional circle with length $L$, the Hamiltonian is
\begin{eqnarray}
H = -\frac{iv}{2}\int_{0}^{L}\mathrm{d}x\sum_{a=1}^{5}\chi^{a}\partial_{x}\chi^{a},
\label{eq:MajorHami}
\end{eqnarray}
where $\chi^{a}(x)$ are Majorana operators with commutation relation $\{\chi^{a}(x),\chi^{b}(x^{\prime})\}=\delta (x-x^{\prime})\delta_{ab}$ and $v$ is the velocity. For antiperiodic (periodic) boundary condition $\chi^{a}(x+L)=-\chi^{a}(x)$ [$\chi^{a}(x+L)=\chi^{a}(x)$], the mode expansion is
\begin{eqnarray}
\chi^{a}(x) = \frac{1}{\sqrt{L}}\sum_{n}\chi_{n}^{a}e^{2\pi inx/L}
\label{eq:ModeExpan}
\end{eqnarray}
with $n\in\mathbb{Z}+1/2$ ($n\in\mathbb{Z}$) and $\{\chi^{a}_{n},\chi^{b}_{m}\} = \delta_{n+m,0}\delta_{ab}$. The antiperiodic (periodic) boundary condition is commonly referred to as the Neveu-Schwarz (Ramond) sector.

The chiral SO(5)$_{1}$ WZW model has three chiral primary fields: the identity field $\boldsymbol{1}$ [SO(5) singlet], the fermion field $\boldsymbol{v}$ [SO(5) vector], and the twist field $\boldsymbol{s}$ [SO(5) spinor]. The states in each Kac-Moody tower can be constructed easily using the Majorana operators~\cite{Francesco1997,Lahtinen2014}. For the Kac-Moody tower associated with the identity field $\boldsymbol{1}$, the primary state is the vacuum $|0\rangle_{\mathrm{NS}}$ of the Neveu-Schwarz (NS) sector, and descendant states are built by acting an \textit{even} number of $\chi^{a}_{-n}$ ($n\in \mathbb{Z}+1/2$ and $n>0$) on $|0\rangle_{\mathrm{NS}}$ [under the constraint of Pauli's exclusion principle $(\chi^{a}_{-n})^{2}=0$]. The counting of states in the identity sector can be seen from the chiral character
\begin{eqnarray}
\mathrm{ch}_{\boldsymbol{1}}(q) & = & \frac{1}{2} q^{-5/48} \left[ \prod^{+\infty}_{n=0} (1+q^{n+1/2})^{5} + \prod^{+\infty}_{n=0} (1-q^{n+1/2})^{5} \right] \nonumber \\
& = & q^{-5/48} \left( 1+10q+30q^{2}+85q^{3}+\cdots \right),
\label{eq:char1}
\end{eqnarray}
where the coefficient of $q^{n-1}$ is the number of states at the $n$-th level. The states at each level form SO(5) representations. The primary state $|0\rangle_{\mathrm{NS}}$ is the only state at the first level, so it is obviously an SO(5) singlet. The ten states at the second level have the form of $\chi^{a}_{-1/2}\chi^{b}_{-1/2}|0\rangle_{\mathrm{NS}}$ ($1 \leq a<b \leq 5$) and they belong to the SO(5) adjoint representation. For higher levels, it is more convenient to label an SO(5) irreducible representation by two integers $p,p^{\prime}$ ($p \geq p^{\prime} \geq 0$), whose dimension is $d_{(p,p^{\prime})}=(1+p^{\prime})(1+p-p^{\prime})(2+p)(3+p+p^{\prime})/6$~\cite{Yang1978,Zhang2001}. Because each mode $\chi_{-n}^{a}$ transforms as an SO(5) vector [the fundamental representation $(1,1)$], the SO(5) representations of all states can be worked out using the tensor product decomposition of SO(5) vectors (see Refs.~\cite{Leung1993,Schuricht2008,TuHH2009} for some useful results; also note that SO(5)$=$Sp(4) at the Lie algebra level). The results in the identity sector are listed in Table~\ref{tab:count1}.

\begin{table*}[tbp]
\begin{ruledtabular}
\begin{tabular}{cccc}
Level & Number of states & SO(5) quantum number & SU(2) quantum number \\
\hline
1 & 1  & $(0,0)$ & $0$  \\
2 & 10 & $(2,0)$ & $0 \oplus 1^{\times 3}$  \\
3 & 30 & $(0,0) \oplus (1,1) \oplus (2,0) \oplus (2,2)$ & $0^{\times 7} \oplus 1^{\times 6} \oplus 2$ \\
4 & 85 & $(0,0) \oplus (1,1) \oplus (2,0)^{\times 3} \oplus (2,2) \oplus (3,1)$ & $0^{\times 11} \oplus 1^{\times 18} \oplus 2^{\times 4}$
\end{tabular}
\end{ruledtabular}
\caption{The number of states and their quantum numbers in the identity sector of the chiral SO(5)$_{1}$ WZW model.}
\label{tab:count1}
\vspace{1em}
\begin{ruledtabular}
\begin{tabular}{cccc}
Level & Number of states & SO(5) quantum number & SU(2) quantum number \\
\hline
1 & 5   & $(1,1)$ & $0^{\times 2} \oplus 1$  \\
2 & 15  & $(1,1) \oplus (2,0)$ & $0^{\times 3} \oplus 1^{\times 4}$  \\
3 & 56  & $(0,0) \oplus (1,1)^{\times 2} \oplus (2,0) \oplus (3,1)$ & $0^{\times 8} \oplus 1^{\times 11} \oplus 2^{\times 3}$ \\
4 & 130 & $(0,0) \oplus (1,1)^{\times 3} \oplus (2,0)^{\times 3} \oplus (2,2) \oplus (3,1)^{\times 2}$ & $0^{\times 17} \oplus 1^{\times 26} \oplus 2^{\times 7}$
\end{tabular}
\end{ruledtabular}
\caption{The number of states and their quantum numbers in the fermion sector of the chiral SO(5)$_{1}$ WZW model.}
\label{tab:count2}
\vspace{1em}
\begin{ruledtabular}
\begin{tabular}{cccc}
Level & Number of states & SO(5) quantum number & SU(2) quantum number \\
\hline
1 & 4   & $(1,0)$ & $\tfrac{1}{2}^{\times 2}$  \\
2 & 20  & $(1,0) \oplus (2,1)$ & $\tfrac{1}{2}^{\times 6} \oplus \tfrac{3}{2}^{\times 2}$  \\
3 & 60  & $(1,0)^{\times 2} \oplus (2,1)^{\times 2} \oplus (3,0)$ & $\tfrac{1}{2}^{\times 14} \oplus \tfrac{3}{2}^{\times 8}$ \\
4 & 160 & $(1,0)^{\times 4} \oplus (2,1)^{\times 4} \oplus (3,0)^{\times 2} \oplus (3,2)$ & $\tfrac{1}{2}^{\times 34} \oplus \tfrac{3}{2}^{\times 20} \oplus \tfrac{5}{2}^{\times 2}$
\end{tabular}
\end{ruledtabular}
\caption{The number of states and their quantum numbers in the twist sector of the chiral SO(5)$_{1}$ WZW model.}
\label{tab:count3}
\end{table*}

The SO(5) symmetry is expected to emerge at low energy in sufficiently large systems. However, the actual states to be investigated only have an exact SU(2) symmetry at the microscopic level. Numerical results shall be analyzed based on the SU(2) symmetry, so we decompose the chiral SO(5)$_{1}$ conformal towers into SU(2) representations. The free Majorana representation in Eq.~(\ref{eq:MajorHami}) is convenient for this purpose, where three of five Majorana fermions (with label $a=1,2,3$) are identified to form an SU(2) triplet ($S=1$), and the remaining two (with label $a=4,5$) transform as SU(2) singlets ($S=0^{\times 2}$)~\footnote{In principle, the five Majorana fermions may transform under the $S=2$ representation of SU(2), but this choice does not agree with our numerical results.}. This makes the SU(2) quantum numbers of the states transparent. For instance, the ten states at the second level are decomposed into the SU(2) representations $S=0\oplus 1^{\times 3}$ (one singlet and three triplets), which can be expressed as $\chi^{4}_{-1/2} \chi^{5}_{-1/2}|0\rangle_{\mathrm{NS}}$, $\chi^{a}_{-1/2}\chi^{b}_{-1/2}|0\rangle_{\mathrm{NS}}$ ($1 \leq a<b \leq 3$), and $\chi^{a}_{-1/2}\chi^{b}_{-1/2}|0\rangle_{\mathrm{NS}}$ ($a=1,2,3$ and $b=4,5$), respectively. Table~\ref{tab:count1} also provides the SU(2) quantum numbers.

For the fermion sector, the states are obtained by acting an \textit{odd} number of $\chi^{a}_{-n}$ ($n\in\mathbb{Z}+1/2$ and $n>0$) on $|0\rangle _{\mathrm{NS}}$, where the primary state $\chi_{-1/2}^{a}|0\rangle_{\mathrm{NS}}$ is an SO(5) vector with conformal weight $h_{\boldsymbol{v}}=1/2$. The chiral character
\begin{eqnarray}
\mathrm{ch}_{\boldsymbol{v}}(q) &=& \frac{1}{2} q^{-5/48} \left[ \prod^{+\infty}_{n=0} (1+q^{n+1/2})^{5} - \prod^{+\infty}_{n=0} (1-q^{n+1/2})^{5} \right] \nonumber \\
&=& q^{-5/48} q^{1/2} \left( 5+15q+56q^{2}+130q^{3}+\cdots \right)
\label{eq:char2}
\end{eqnarray}
tells us the counting of states. For the twist sector $\boldsymbol{s}$, the states reside in the Ramond (R) sector and there is a subtlety due to the fermion zero modes. The chiral primary states $|s\rangle_{\mathrm{R}}$ have four components (with conformal weight $h_{\boldsymbol{s}}=5/16$) and transform as an SO(5) spinor. It is denoted by $(1,0)$ and identified to have SU(2) quantum number $S=\frac{1}{2}\oplus\frac{1}{2}$. The descendant states can be constructed by acting a number of $\chi^{a}_{-n}$ on $|s\rangle_{\rm R}$ ($n\in\mathbb{Z}$ and $n>0$). The chiral character
\begin{eqnarray}
\mathrm{ch}_{\boldsymbol{s}}(q) &=& \frac{1}{8} q^{5/24} \prod^{\infty}_{n=0} (1+q^{n})^{5} \nonumber \\
& = & q^{-5/48} q^{5/16} \left( 4+20q+60q^{2}+160q^{3}+\cdots \right)
\end{eqnarray}
tells us the counting of states. The results are presented in Tables~\ref{tab:count2} and~\ref{tab:count3}.

\section{Chiral Spin Liquid}
\label{liquid}

\begin{figure}
\centering
\includegraphics[width=0.46\textwidth]{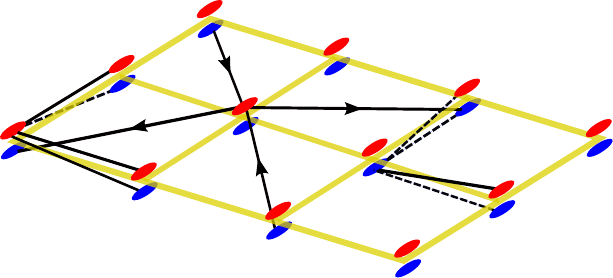}
\caption{Schematics of the spinless Chern insulator model with $C=2$. The two states $A,B$ on each site are represented by red and blue dots. The nearest-neighbor hopping amplitudes all have absolute value $1$ and the sign is $+1$ ($-1$) along the solid (dashed) line. The next-nearest-neighbor hopping amplitude is $i/\sqrt{2}$ ($-i/\sqrt{2}$) along (against) the arrow.}
\label{Figure1}
\end{figure}

In the continuum setting, the parton construction usually generates states that populate mutiple Landau levels. If lowest Landau level projection is not performed, the states are somewhat analytically tractable but not really convenient for numerical purposes~\cite{Bandyo2020,Anand2021}. To this end, we consider a lattice model of free fermions whose Bloch bands possess non-zero Chern numbers~\cite{ZhangY2013,ZhangY2014}. As illustrated in Fig.~\ref{Figure1}, the model is defined on the square lattice with two orbitals per site. The creation (annihilation) operators for the fermions are denoted as $c^{\dag}_{j,u}$ ($c_{j,u}$) where $j$ labels the lattice site and $u=A,B$ distinguishes the two orbitals on each site. The Hamiltonian is
\begin{align}
H_{\rm CI} & = \sum_{\langle jk \rangle} \left[ c^{\dag}_{j,A} c_{k,A} + {\rm H.c.} \right] + \sum_{\langle jk \rangle} (-1) \left[ c^{\dag}_{j,B} c_{k,B} + {\rm H.c.} \right] \nonumber \\
&  + \sum_{\langle jk \rangle_{x}} \left[ c^{\dag}_{j,A} c_{k,B} + {\rm H.c.} \right] + \sum_{\langle jk \rangle_{y}} (-1) \left[ c^{\dag}_{j,B} c_{k,A} + {\rm H.c.} \right] \nonumber \\
&  + \sum_{\langle\langle jk \rangle\rangle} (-i/\sqrt{2}) \left[ c^{\dag}_{j,A} c_{k,B} + {\rm H.c.} \right] \nonumber \\
& + \sum_{\langle\langle jk \rangle\rangle} (i/\sqrt{2}) \left[ c^{\dag}_{j,B} c_{k,A} + {\rm H.c.} \right],
\label{eq:PartonHami}
\end{align}
where $\langle jk \rangle$ denotes nearest neighbors, $\langle jk \rangle_{x,y}$ denotes nearest neighbors along the $x$ or $y$ direction, and $\langle\langle jk \rangle\rangle$ denotes next nearest neighbors. The number of unit cells along the two directions are denoted as $N_{x}$ and $N_{y}$. $H_{\rm CI}$ can be diagonalized to yield the single-particle energy levels $\varepsilon_{i}$ and the creation (annihilation) operators $\Gamma^{\dag}_{i}$ ($\Gamma_{i}$) for the single-particle eigenstates. If the system has periodic boundary conditions along both directions, we get two bands with Chern number $C={\pm}2$. A lattice counterpart of the $\nu=2$ IQH state in Landau levels is obtained when the lower band is completely filled. The fermionic operators are appended with another index $\alpha$ that has two values called spin-up and spin-down. The fermions are assembled to form the spin-1/2 operators
\begin{eqnarray}
S^{a}_{j,u} = \frac{1}{2} \sum_{\alpha,\beta=\uparrow,\downarrow} c^{\dag}_{j,u,\alpha} \tau^{a}_{\alpha\beta} c_{j,u,\beta},
\label{eq:Spin}
\end{eqnarray}
where $a=x,y,z$ and $\tau^{a}$ are the Pauli matrices. A many-body state of the fermions is also a valid state for the spins if each orbital is singly occupied (i.e. $\sum_{\alpha=\uparrow,\downarrow} c^{\dag}_{j,u,\alpha} c_{j,u,\alpha}=1$).

In the spin language, the parton FQH state turns into a chiral spin liquid. Roughly speaking, the lower bands of fermions with both spins are completely filled, the two product states are combined, and Gutzwiller projection $\mathcal{P}_{\rm G}$ is applied to ensure single occupancy of the orbitals. The tensor network methods are employed to find the MPS representation of the chiral spin liquid~\cite{WuYH2020,JinHK2020,Aghaei2020,Petrica2021,JinHK2022}. The system shall be placed on a cylinder (which is periodic along only one direction) instead of a torus. This is largely due to the fact that MPS is not very efficient for periodic systems. While topological properties can be extracted easily from the MES, their construction is not always transparent. For the torus case, a well-established prescription (although its validity has not been proved in general) is using both periodic and anti-periodic boundary conditions for the partons~\cite{ZhangY2013,ZhangY2014,LiuZX2018}. The problem could be even more tricky on the cylinder (with the $x$ direction being open)~\cite{TuHH2013,WuYH2020}. For each member of the MES, the cylinder is threaded by a flux line that connects two dangling anyons localized at its ends. Based on previous experience, we expect that MES can be constructed only if the partons have single-particle edge states at exactly zero energy. The anyons should form a state with SU(2) symmetry such that the edge states are labeled by proper quantum numbers. A more ambitious goal would be to elevate the symmetry to SO(5) explicitly. One should also keep in mind that two different states of free partons may turn out to be the same after Gutzwiller projection.

\begin{figure}
\centering
\includegraphics[width=0.48\textwidth]{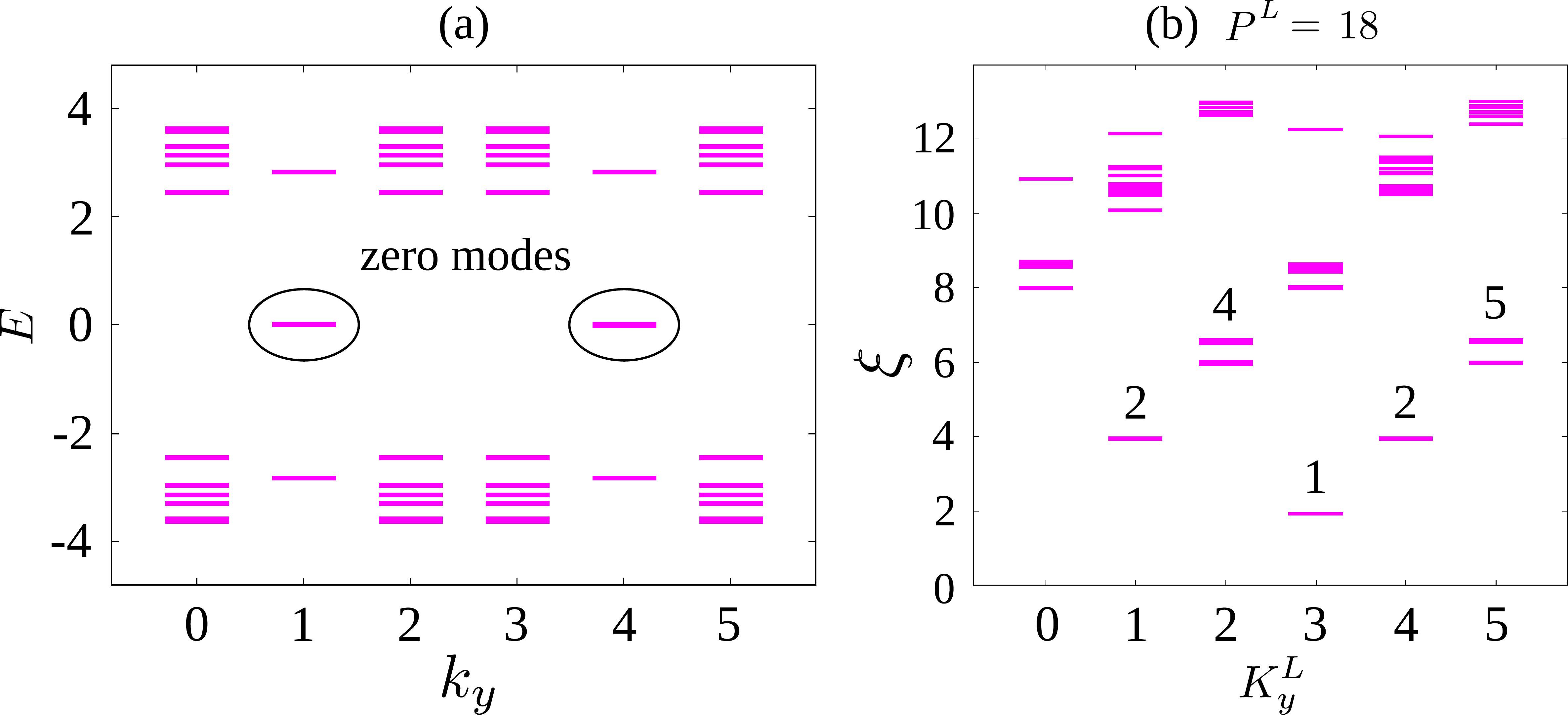}
\caption{(a) The single-particle energy spectrum of the Hamiltonian $H_{\rm CI}$ on a cylinder with $N_{x}=6$ and $N_{y}=6$. (b) The entanglement spectrum of the free fermion state $|\Psi_{\rm CI}\rangle$ with $N_{x}=6$ and $N_{y}=6$ (see text for its definition).}
\label{Figure2}
\end{figure}

\begin{figure*}
\centering
\includegraphics[width=0.95\textwidth]{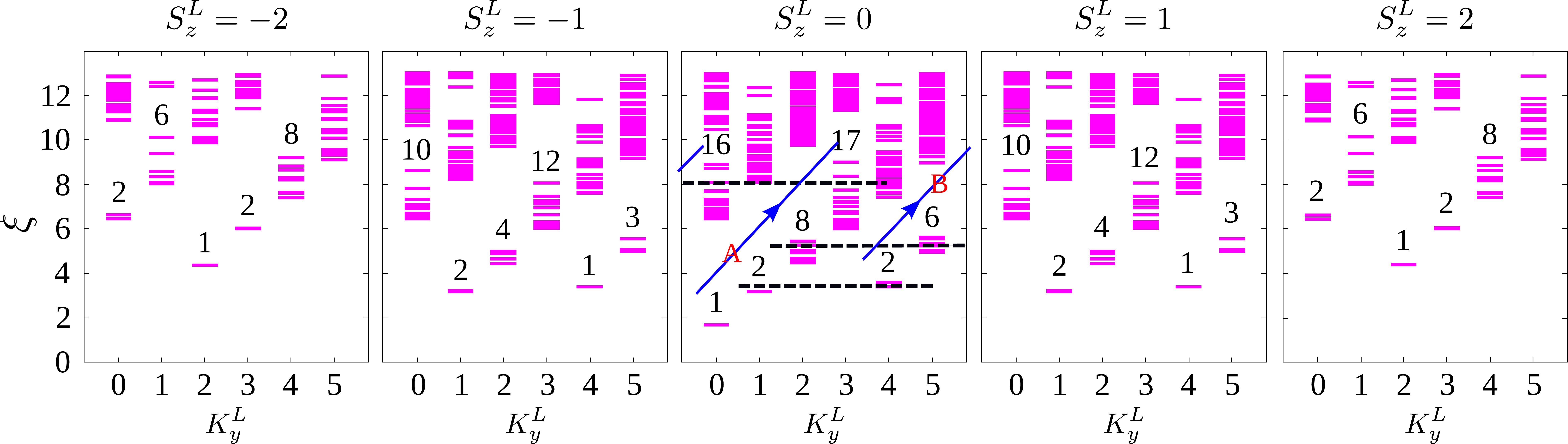}
\caption{The entanglement spectrum of the chiral spin liquid with $N_{x}=6$ and $N_{y}=6$ in the identity sector. The counting of levels is shown in the panels. Three towers of CFT excited states are indicated using dashed lines in the middle panel. Two branches of levels are marked by blue arrows in the middle panel. The sixteen levels at $K^{L}_{y}=0$ belongs to the $B$ branch due to periodicity of momentum.}
\label{Figure3}
\end{figure*}

\begin{figure*}
\centering
\includegraphics[width=0.95\textwidth]{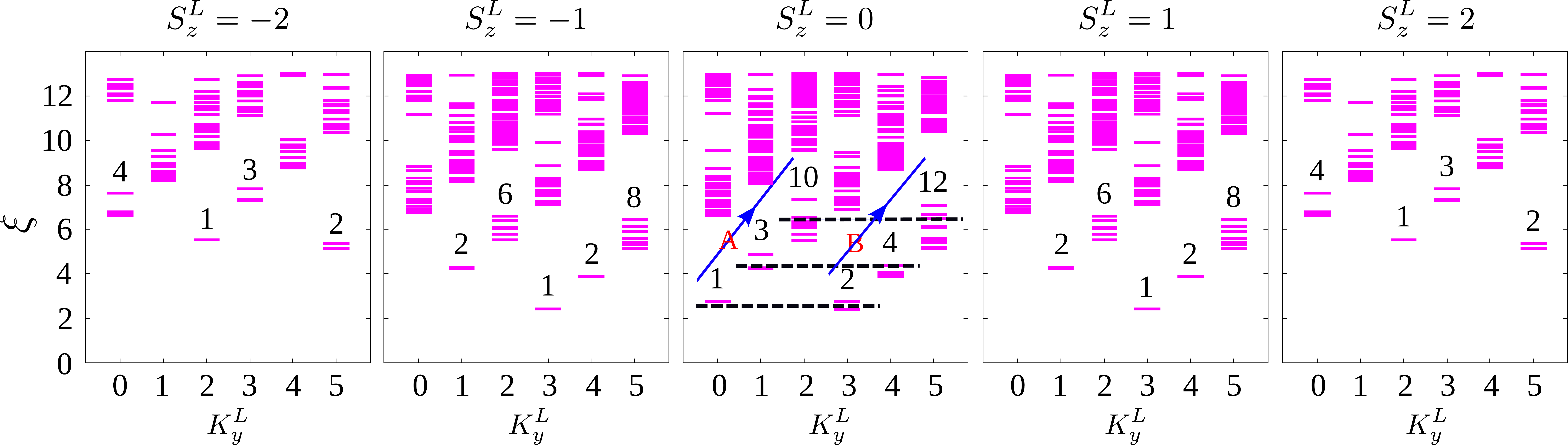}
\caption{The entanglement spectrum of the chiral spin liquid with $N_{x}=6$ and $N_{y}=6$ in the fermion sector. The notation is the same as in Fig.~\ref{Figure3}.}
\label{Figure4}
\end{figure*}

\begin{figure*}
\centering
\includegraphics[width=0.95\textwidth]{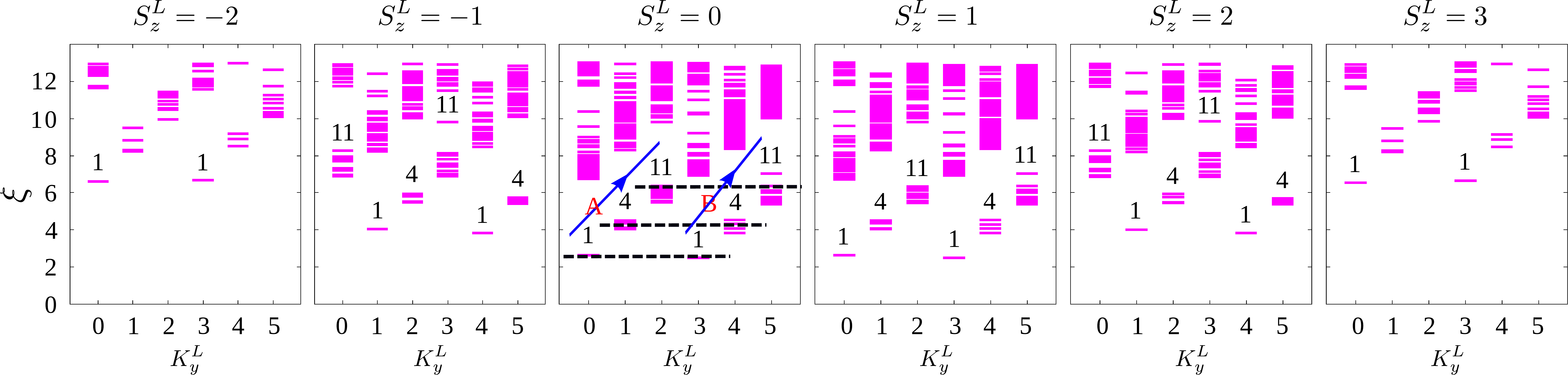}
\caption{The entanglement spectrum of the chiral spin liquid with $N_{x}=6$ and $N_{y}=6$ in the twist sector. The notation is the same as in Fig.~\ref{Figure3}.}
\label{Figure5}
\end{figure*}

$H_{\rm CI}$ contains four zero modes (two per edge) in two types of systems: 1) the $y$ direction is periodic and $N_{y}$ is a multiple of four; 2) the $y$ direction is antiperiodic and $N_{y}$ is a multiple of two (but not four). To be specific, we focus on the system with $N_{x}=6$ and $N_{y}=6$. Its single-particle spectrum is presented in Fig.~\ref{Figure2} (a) where $k_{y}$ is the momentum along the $y$ direction (in units of $2\pi/N_{y}$). The creation operators for the zero modes are denoted as $\zeta^{\dag}_{L,k_{y},\alpha}$ and $\zeta^{\dag}_{R,k_{y},\alpha}$, where $L$ and $R$ indicate the two ends, $k_{y}=1,4$, and $\alpha=\uparrow,\downarrow$. The negative energy states of both spins are fully occupied to generate the state
\begin{eqnarray}
|\Phi\rangle = \prod_{\varepsilon_{i}<0} \prod_{\alpha=\uparrow,\downarrow} \Gamma^{\dag}_{i,\alpha} |0\rangle
\label{eq:PhiState}
\end{eqnarray}
and the zero modes are populated in certain combinations to generate the MES. When one zero mode is occupied at each boundary of the cylinder, the system transforms under the SU(2) reducible representation $\frac{1}{2}\oplus\frac{1}{2}$, which corresponds to the SO(5) spinor representation. It can be argued that each zero mode is associated with a twist anyon of the SO(5)$_{1}$ theory, and the occupation of multiple zero modes at the same boundary can be viewed as fusion of anyons. This picture is useful in finding all topological sectors and will be justified later by numerical results. The identity sector has trivial anyons so partons in the zero modes at both ends form SO(5) singlet states. If all zero modes on one end are empty, this end obviously hosts no anyon. However, we cannot leave both ends empty because this would not provide us a valid spin state (i.e., the single occupancy of each orbital cannot be fulfilled). Another way that generates an SO(5) singlet is to populate all zero modes on one end. This analysis leads to two states
\begin{eqnarray}
&& |\Psi_{\boldsymbol{1},a}\rangle = \mathcal{P}_{\rm G} \; \zeta^{\dag}_{L,1,\uparrow} \zeta^{\dag}_{L,1,\downarrow} \zeta^{\dag}_{L,4,\uparrow} \zeta^{\dag}_{L,4,\downarrow} |\Phi\rangle, \nonumber \\
&& |\Psi_{\boldsymbol{1},b}\rangle = \mathcal{P}_{\rm G} \; \zeta^{\dag}_{R,1,\uparrow} \zeta^{\dag}_{R,1,\downarrow} \zeta^{\dag}_{R,4,\uparrow} \zeta^{\dag}_{R,4,\downarrow} |\Phi\rangle.
\label{eq:PsiStateI}
\end{eqnarray}
The overlap between them is $0.9265$ and we believe that they actually represent the same state~\footnote{The overlap is limited by the bond dimension of the MPS. For the system with $N_{x}=4$ and $N_{y}=4$, similar states have been constructed with higher accuracy and their overlap is $0.9997$. It is almost certain that the overlap in the case with $N_{x}=6$ and $N_{y}=6$ could be further improved.}. The next step is to construct the fermion sector for which each end hosts an Abelian anyon and the system realizes one of the SO(5) vector states. This can be achieved if both ends have two partons form certain superpositions in the zero modes. We have constructed four states
\begin{eqnarray}
&& |\Psi_{\boldsymbol{v},a}\rangle = \mathcal{P}_{\rm G} \; \zeta^{\dag}_{L,1,\uparrow} \zeta^{\dag}_{L,1,\downarrow} \zeta^{\dag}_{R,1,\uparrow} \zeta^{\dag}_{R,1,\downarrow} |\Phi\rangle, \nonumber \\
&& |\Psi_{\boldsymbol{v},b}\rangle = \mathcal{P}_{\rm G} \; \zeta^{\dag}_{L,1,\uparrow} \zeta^{\dag}_{L,1,\downarrow} \zeta^{\dag}_{R,4,\uparrow} \zeta^{\dag}_{R,4,\downarrow} |\Phi\rangle, \nonumber \\
&& |\Psi_{\boldsymbol{v},c}\rangle = \mathcal{P}_{\rm G} \; \zeta^{\dag}_{L,4,\uparrow} \zeta^{\dag}_{L,4,\downarrow} \zeta^{\dag}_{R,1,\uparrow} \zeta^{\dag}_{R,1,\downarrow} |\Phi\rangle, \nonumber \\
&& |\Psi_{\boldsymbol{v},d}\rangle = \mathcal{P}_{\rm G} \; \zeta^{\dag}_{L,4,\uparrow} \zeta^{\dag}_{L,4,\downarrow} \zeta^{\dag}_{R,4,\uparrow} \zeta^{\dag}_{R,4,\downarrow} |\Phi\rangle
\label{eq:PsiStateV}
\end{eqnarray}
and found that their overlaps are all close to $1$. This means that they represent the same state and their is no need to form superpositions. For the identity (fermion) sector, four (two) zero modes on one end are occupied and none (two) zero modes on the other end is occupied. It makes us conjecture that the two ends in the twist sector have three zero modes and one zero mode occupied, respectively. This intuition is supported by the fact that proper superpositions of three zero modes can form the SO(5) spinor. However, we shall use the relatively simple state
\begin{eqnarray}
&& |\Psi_{\boldsymbol{s}}\rangle = \mathcal{P}_{\rm G} \; \zeta^{\dag}_{L,1,\uparrow} \zeta^{\dag}_{L,4,\uparrow} \zeta^{\dag}_{L,4,\downarrow} \zeta^{\dag}_{R,1,\downarrow} |\Phi\rangle
\label{eq:PsiStateS}
\end{eqnarray}
rather than any superpositions (see below for more discussion).

The edge states can be probed using entanglement spectrum~\cite{LiH2008}. The cylinder is divided into two parts and the reduced density matrix (RDM) of the left half is computed. This operation creates a virtual edge on the cylinder and expose the edge physics via the spectrum of the RDM~\cite{QiXL2012}. In practice, the Schmidt decomposition of the many-body states is sufficient because the Schmidt values are simply the square roots of the eigenvalues of the RDM. The Schmidt states are labeled by three quantum numbers: the total spin $S^{L}(S^{L}+1)$, the $z$-component $S^{L}_{z}$ of total spin, and the momentum $K^{L}_{y}$ along the $y$ direction. The logarithms of the eigenvalues are plotted versus $S^{L}_{z}$ and $K^{L}_{y}$ to generate the entanglement spectrum. Let us first explain the identity sector shown in Fig.~\ref{Figure3}. We do not need to compute the total spin explicitly because it can be deduced by comparing the spectrum in different $S^{L}_{z}$ sectors. One can see a unique lowest level and 10 quasi-degenerate second levels (2+1+2+2+2+1). The third and fourth levels are more difficult to identify because the number increases rapidly and the degeneracy is not very good. Nevertheless, it is reasonable to claim that the number of third levels is 30 (1+4+3+8+6+4+3+1) and that of fourth levels is 85 (2+2+10+12+16+17+10+12+2+2). The lowest level appears in the $S^{L}_{z}=0$ sector, which implies that it has total spin $S^{L}=0$. Among the ten second levels, four of them appear in the $S^{L}_{z}=0$ sector and three of them appear in each of the $S^{L}_{z}={\pm}1$ sectors. This means that the second levels can be grouped as one $S^{L}=0$ and three $S^{L}=1$. The same analysis can be performed for the third and fourth levels. The fermion sector shown in Fig.~\ref{Figure4} can be analyzed in the same way because the state is also an SU(2) singlet. In total, the results clearly confirm the theoretical predictions summarized in Tables~\ref{tab:count1} and~\ref{tab:count2}. The twist sector is more complicated. While SU(2) or SO(5) symmetric states can be constructed, it is actually undesirable to do that~\cite{WuYH2020}. If either symmetry is fulfilled, the entanglement spectrum would not exhibit pure SO(5) towers but multiple copies of them. This problem is resolved if the anyons are polarized to one component of the superposition as in Eq.~(\ref{eq:PsiStateS}). The price to pay is that the entanglement levels do not have definite total spin values, but this does not invalidate our theoretical prediction. As one can see from Fig.~\ref{Figure5}, it is still feasible to group the levels in a manner that is consistent with the SO(5) towers in Table~\ref{tab:count3}.

An important feature of the entanglement spectrum that was not revealed by our analysis in Sec.~\ref{edge} is the existence of multiple branches. More specifically, the levels in the $S^{L}_{z}=0$ sector can be organized as two branches marked by the arrows in Fig.~\ref{Figure3}. One branch contains 2 second levels and 8 third levels, while the other branch contains 2 second levels and 6 third levels. This property can be traced back to the edge states of the parton Chern insulator. It is sufficient to consider only one spin since the two components are decoupled. The spin-up (or spin-down) parton state in which both zero modes on the left are occupied is denoted as $|\Psi_{\rm CI}\rangle$. As shown in Fig.~\ref{Figure2} (b), the entanglement spectrum of $|\Psi_{\rm CI}\rangle$ also has multiple branches (labeled by the number of fermions $P^{L}$ and the momentum $K^{L}_{y}$ along the $y$ direction). In fact, this seems to be a quite general feature in lattice models with $C>1$~\cite{WangYF2012,Trescher2012,YangS2012}. The entanglement spectrum of the chiral spin liquid is obtained by combining the entanglement spectrum of the spin-up and spin-down partons. The momenta of two levels can be added directly because Gutzwiller projection does not change momentum. For example, two copies of the lowest level with $K^{L}_{y}=3$ in Fig.~\ref{Figure2} (b) generate the lowest level with $K^{L}_{y}=3+3 \equiv 0$ in Fig.~\ref{Figure3}. The excited levels are more complicated because it is very likely that the combined levels are not all linearly independent. Nevertheless, the 4 second levels in the $S^{L}_{z}=0$ sector of Fig.~\ref{Figure3} are easy to construct. We can combine the two levels with $K^{L}_{y}=1$ ($K^{L}_{y}=4$) and the lowest level with $K^{L}_{y}=3$ in Fig.~\ref{Figure2} (b) to produce the two levels with $K^{L}_{y}=1+3=4$ ($K^{L}_{y}=4+3 \equiv 1$) in Fig.~\ref{Figure3}.

\section{Conclusions}
\label{conclusion}

In summary, we have uncovered that the edge state of a non-Abelian topological state is described by SO(5)$_{1}$ CFT. The state has been studied previously in different disguises, but the {\it emergent} SO(5) symmetry was never pointed out to the best of our knowledge. This is reminiscent of the fact that the $\nu=1/2$ bosonic Laughlin state has an emergent SU(2) symmetry~\cite{Kalmeyer1987} and the bilayer Halperin $221$ state has an emergent SU(3) symmetry~\cite{Ardonne1999}. The SU(2) gauge field in the parton description and the emergent SO(5) symmetry also distinguish the present state from other SO(5)$_{1}$ chiral spin liquids that were constructed by coupling topological superconductors to $\mathbb{Z}_{2}$ gauge fields (in the spirit of Kitaev's 16-fold way~\cite{Kitaev2006-2}) and mostly have exact SO(5) symmetry at the microscopic level~\cite{TuHH2013-2,WangJC2019,Chulli2020,ZhangHC2022}. The counting of edge states is computed using CFT techniques and confirmed in the entanglement spectrum.

The modular matrices of the SO(5)$_{1}$ chiral spin liquid have been computed using Monte Carlo methods~\cite{ZhangY2013,ZhangY2014}. It would be very useful if the MPS framework can provide more accurate results. In principle, this can be done using the MPS representation of the chiral spin liquid on infinte cylinders~\cite{Cincio2013}. Technical challenges are expected in the implementation and computational time may be substantial. Another obvious extension for future works is to consider parton bands with Chern number $C>2$. The edge theory of such systems is expected to be the U(2$C$)$_{1}$/SU(2)$_{C}$ CFT, which is actually equivalent to the Sp(2$C$)$_{1}$ CFT based on a duality relation~\cite{Aharony2017}. The $C=2$ case studied in the present work also fits into this framework since Sp(4)$_{1}$=SO(5)$_{1}$. Gutzwiller wave functions for Sp(2$C$)$_{1}$ chiral spin liquids with $C>2$ and possible microscopic models that can realize them are interesting topics that deserve further investigations.

\vspace{1em}

{\it Note added} --- Recently we notice a paper that has some overlap with this work~\cite{Anand2022}. 

\vspace{1em}

\section*{Acknowledgements}

We are grateful to Meng Cheng for pointing out the duality relation and Hua-Chen Zhang for helpful discussions on SO(5)$_{1}$ CFT. Y.-H. W. is supported by the NNSF of China under grant No. 12174130. H.-H. T. is supported by the Deutsche Forschungsgemeinschaft through project A06 of SFB 1143 under project No. 247310070.

\bibliography{../ReferCollect}

\end{document}